\documentclass[twocolumn,english,aps,prb,showpacs,floats,floatfix]{revtex4}
\usepackage[T1]{fontenc}
\usepackage[latin1]{inputenc}
\usepackage{amsmath}
\usepackage{graphicx}
\usepackage{amssymb}

\makeatletter
\usepackage{graphicx}
\usepackage{psfrag}
\usepackage{graphicx}
\providecommand{\LyX}{L\kern-.1667em\lower.25em\hbox{Y}\kern-.125emX\@}

\usepackage{babel}
\makeatother
\begin{document}

\title{Dynamical Mean Field Theory of Quantum Stripe Glasses}

\author{Harry Westfahl Jr.$^{1,2}$, J\"{o}rg Schmalian$^{2}$, and Peter
G Wolynes$^{3}$.}

\affiliation{$^{1}$Laboratório Nacional de Luz Síncrotron - ABTLuS, Campinas,
SP 13084-971, BRAZIL}

\affiliation{$^{2}$Department of Physics and Astronomy and Ames Laboratory, Iowa
State University, Ames, IA 50011}

\affiliation{$^{3}$Department of Chemistry and Biochemistry, University of California,
San Diego, La Jolla, CA 92093 }

\date{\today{}}


\begin{abstract}
We present a many body approach for non-equilibrium behavior and self-generated
glassiness in strongly correlated quantum systems. It combines the
dynamical mean field theory of equilibrium systems with the replica
theory for classical glasses without quenched disorder. We apply this
approach to study a quantized version of the Brazovskii model and
find a self-generated quantum glass that remains in a quantum mechanically
mixed state as $T\rightarrow0$. This quantum glass is formed by a
large number of competing states spread over an energy region which
is determined within our theory.
\end{abstract}
\maketitle

\section{Introduction}

The formation of glasses upon cooling is a well known phenomenon for
classical liquids. Even without quenched disorder, the relaxation
times become so large that a frozen non-ergodic state is reached before
the nucleation into the crystalline solid sets in. The nucleation
is especially easy to avoid if the material has many polymorphisms
as does for example SiO$_{2}$. The system becomes unable to reach
an equilibrium configuration on the laboratory time scale and exhibits
aging and memory effects. While extrinsic disorder is widely acknowledged
to lead to glassy phenomena in quantum systems, can a similar self
generated glassiness occur for quantum liquids? Candidate materials
for this behavior are strongly correlated electron systems which often
exhibit a competition between numerous locally ordered states with
comparable energies. Examples for such behavior are colossal magneto-resistance
materials,\cite{mang01,Millis96,manganites} cuprate superconductors\cite{nmr01,nmr02,nmr03,nmr3b,nmr04,msr01,msr02,msr03}
and likely low density electron systems\cite{AKS01} possibly close
to Wigner crystallization.

The possibility of self generated quantum glassiness in such systems
and the nature of the slow quantum dynamics has been little explored.
Within a purely classical theory a {}``stripe glass'' state was
recently proposed in Refs.\cite{SW00,SWW00}. This proposal was stimulated
by nuclear magnetic resonance\cite{nmr01,nmr02,nmr03,nmr3b,nmr04}
and $\mu$-spin relaxation experiments\cite{msr01,msr02,msr03} who
found static or quasi-static charge and spin configurations similar
to glassy or disordered systems. A prominent effect which reflects
this slowing down was the {}``wipe-out'' effect where below a certain
temperature the typical time scales become so long that the resulting
rapid spin lattice relaxation cannot be observed anymore\cite{nmr02,nmr03,nmr04}.
More recently, the dynamical behavior of the stripe glass of Refs.\cite{SW00,SWW00},
as determined by Grousson \emph{et al.}\cite{Grousson} was found
to be in quantitative agreement with NMR-experiments\cite{nmr05}.
Finally, recent $\mu$-spin experiments\cite{msr02,msr03} analyzed
the freezing temperatures as function of charge carrier concentration
and disorder concentration and found a quantum glass transition which
was insensitive to the amount of disorder added to the system. This
latter experiment, which seems to support a generic explanation for
glassiness,not caused by impurities, also demonstrated the need for
a more detailed investigation of the quantum regime of glassy systems.
It is then important to develop appropriate theoretical tools to predict
whether a given theoretical model for a quantum many body system will
exhibit self-generated glassiness where the system forms a glass for
arbitrary weak disorder.

In this paper we develop a general approach to self generated quantum
glasses that combines the dynamical mean field theory (DMFT) of quantum
many body systems\cite{dmftreview1,dmftreview2,dmftreview3} with
the replica technique of Refs.\onlinecite{Mon95,MP991} which describes
classical glasses without quenched disorder. In addition, we also
go beyond strict mean field theory and estimate the spectrum of competing
quantum states after very long times. This approach is applied to
study a model with competing interactions which has a fluctuation
induced first-order transition to a striped (2D) / lamellar (3D) phase.
A quantum glass is shown to be in a quantum mechanical mixed state
even at $T=0$, formed by a large number of states which can be considerably
above the true ground state. Using the concept of an \emph{effective
temperature}\cite{Tool46,N98,CK93} for the distribution of competing
ground states, our theory allows the investigation of glassy non-equilibrium
dynamics in quantum systems using standard techniques of equilibrium
quantum many body theory. On the mean field level, our theory is in
complete agreement with the explicit dynamical non-equilibrium approach
developed by Cugliandolo and Lozano\cite{CL99} for quenched disordered
spin glasses with entropy crisis, which includes weak long-term memory
effects and the subtle interplay of aging and stationary dynamics.

We find that when quantum fluctuations are weak, i.e., for a small
quantum parameter, $c$ (for a definition see Eq.\ref{qbraz} below)
, the glass transition resembles that of classical models\cite{KTW89}
that exhibit a dynamical transition at a temperature $T_{\text{A}}$
(where mean field theory becomes non-ergodic) and a Kauzmann entropy
crisis at $T_{\text{K}}<T_{\text{A}}$ if equilibrium were to be achieved.
The actual laboratory glass transition is located between these two
temperatures and depends for example on the cooling rate of the system.
Beyond a critical $c$, $T_{\text{K}}$ and $T_{\text{A}}$ merge
and the transitions change character. Within mean field theory, a
discontinuous change of the relevant quantum mechanical states occurs
at the glass transition. Even going beyond mean field, the system
remains in a mixed quantum state however with a non-extensive number
of relevant quantum states. Still, at the quantum glass transition
a discontinuous change of the density matrix to a usual quantum liquid
occurs. Put another way, a glass in contact with a bath at $T=0$
is essentially a classical object, qualitatively distinct from a quantum
fluid, enforcing the quantum glass transition to be discontinuous.
Our results clearly support the later scenario as can be seen in Fig.\ref{Fofc}
below.

In classical glass forming liquids an excess or configurational entropy
with respect to the solid state due to an exponentially large number
of metastable configurations emerges below $T_{\text{A}}$. Obviously,
in the quantum limit a glass transition must be qualitatively different.
Even a very large number of long lived excited states cannot compensate
their vanishing Boltzmann weight at equilibrium for $T=0$. An exponentially
large ground state degeneracy on the other hand is typically lifted
by hybridization, saving kinetic energy. In Ref.\onlinecite{KFI99}
it was then argued that for a glassy quantum system the Edwards-Anderson
order parameter vanishes continuously at the quantum glass transition
(in contrast to the classical behavior). On the other hand, in Ref.\onlinecite{Cugl}
it was concluded that in certain spin glasses with quenched disorder
the glass transition at sufficiently small $T$ is of first order.

The outline of this paper is as follows. In section II we introduce
a quantized version of the Brazovskii model of micro-phase separation.
We then discuss the replica approach used in this paper as well as
the dynamical mean field theory employed for the solution of the problem
in sections III and IV, respectively. Conclusions which are based
on the approach developed here but which go beyond the strict mean
field limit are discussed in section V. Finally we present a summary
of our results in the concluding section VI. The equivalence of the
{}``cloned-liquid'' replica approach and the Schwinger-Keldysh theory
of non-equilibrium quantum systems is presented in the appendix.

\section{The model}

We consider a Bose system with field, $\varphi_{\mathbf{x}}$, governed
by the action $\mathcal{S}\left[\varphi_{\mathbf{x}}\right]$ with
competing interactions which cause a glass transition in the classical
limit. Specifically we consider a system with action \begin{eqnarray}
\mathcal{S} & = & \frac{1}{2}\int d^{3}x\int d\tau\left(\left(\frac{\partial\varphi_{\mathbf{x}}}{\partial c\tau}\right)^{2}+r_{0}\varphi_{\mathbf{x}}^{2}+\frac{u}{2}\varphi_{\mathbf{x}}^{4}\right.\label{qbraz}\\
 &  & \left.+2q_{0}^{-2}\left(\left[\nabla^{2}+q_{0}^{2}\right]\varphi_{\mathbf{x}}\right)^{2}\right).\nonumber \end{eqnarray}
 Here, $q_{0}$ is a wave number which supports strong fluctuations
for momenta with amplitude $\left|\mathbf{q}\right|=q_{0}$, i.e.
modulated field configurations. The classical version of the model,
Eq.\ref{qbraz}, was shown by Brazovskii\cite{Braz} to give rise
to a fluctuation induced first order transition to a lamellar or smectic
state. Within equilibrium statistical mechanics, this ordered state
gives the lowest known free energy. In Refs.\onlinecite{SW00,SWW00}
we demonstrated that, within non-equilibrium classical statistical
mechanics, an alternative scenario is a self generated glass. Instead
of the transition to a smectic state, metastable solutions built by
a superposition of large amplitude waves of wavenumber $q_{0}$, but
with random orientations and phases emerge\cite{WWJW}. Those form
the stripe glass state discussed in Ref. \onlinecite{SW00,SWW00}.
It is unclear so far whether the glassy solution occurs only if the
ordered phase can be avoided by super-cooling or whether there is
a parameter regime where the glass is favored regardless of the cooling
rate.

In Eq.\ref{qbraz} we consider additional quantum fluctuations, characterized
by the velocity, $c$. Clearly, quantum fluctuations reduce the tendency
towards a fluctuation induced first order transition. This can be
seen by evaluating the $\left\langle \varphi^{2}\left(x,\tau\right)\right\rangle $
within the spherical approximation. In the classical limit $\left\langle \varphi^{2}\left(x,\tau\right)\right\rangle \sim Tq_{0}^{2}r^{-1/2}$
with renormalized mass $r=r_{0}+u\left\langle \varphi^{2}\left(x,\tau\right)\right\rangle $
and the solution $r=0$ clearly does not exist. The same fluctuations
which suppress the occurrence of a second order transition lead to
a first order transition at the temperature where $r_{0}\simeq uTq_{0}^{2}r_{0}^{-1/2}$.\cite{Braz}
In the quantum limit the behavior is conceptually similar but fluctuations
grow only logarithmically, $\left\langle \varphi^{2}\left(x,\tau\right)\right\rangle \sim q_{0}^{2}\log\left(\frac{\Lambda}{r}\right)$.
For exponentially large correlation length $r^{-1/2}$ there should
also be a \ fluctuation induced first order transition to a smectic,
which might be related to the state proposed in Ref.\onlinecite{FKE}
in the context of strongly correlated quantum systems. Another option
however is the emergence of a stripe glass, even for large quantum
fluctuations, which results in an amorphous modulated state instead.
The investigation of this option will be the specific application
of our theory. Before we go into specifics of the model, Eq.\ref{qbraz},
we develop a general framework for the description of self generated
quantum glasses.

\section{The {}``cloned liquid'' - replica approach}

Competing interactions of a glassy system cause the ground state energy
as well as the excitations to be very sensitive to small additional
perturbations. In order to quantify this we introduce, following Ref.\onlinecite{Mon95},
a static {}``ergodicity breaking'' field $\psi$: \[
\mathcal{S}_{\psi}\left[\varphi\right]=\,\,\mathcal{S}\left[\varphi\right]+\frac{g}{2}\int d\tau d^{3}x\left(\psi_{\mathbf{x}}-\varphi_{\mathbf{x}}\left(\tau\right)\right)^{2}\]
 and take the limit $g\rightarrow0$ eventually. The coupling between
$\varphi$ and $\psi$ will bias the original energy landscape in
the {}``direction'' of the configuration $\psi$, enabling us to
count distinct configurations. Adopting a mean-field strategy, we
assume that even in the quantum limit $\psi$ should be chosen as
static variable, probing only time averaged configurations.

Introducing $\widetilde{f}_{\psi}=-T\log Z_{\psi}$, with the biased
partition function $Z_{\psi}\,\,=\int D\varphi e^{-\mathcal{S}_{\psi}}$,
$\widetilde{f}_{\psi}(T\rightarrow0)$ corresponds to the ground state
energy for a given $\psi$. If there are many competing ground states,
it is natural to assume that $\widetilde{f}_{\psi}$ determines the
probability, $p_{\psi}$, for a given configuration. If we identify
the actual state of the system we gain the information $S_{c}=-\lim_{g\rightarrow0}\int D\psi\, p_{\psi}\log p_{\psi}$.
Maximizing this configurational entropy $S_{c}$ with respect to $p_{\psi}$
yields the usual result \begin{equation}
p_{\psi}\propto\exp\left(-\widetilde{f}_{\psi}/T_{\text{eff}}\right),\label{p}\end{equation}
 where the effective temperature, $T_{\text{eff}}$ is the Lagrange
multiplier enforcing the constraint that the typical energy is $\widetilde{F}=\lim_{g\rightarrow0}\int D\psi\, p_{\psi}\widetilde{f}_{\psi}$.
$T_{\text{eff}}$ is a measure for the width of the energy region
within which the relevant ground state energies can be found. If the
system is glassy, $T_{\text{eff}}>T$ and a quantum mechanically mixed
state results even as $T\rightarrow0$. Thus, a glass will not be
in a pure quantum mechanical state (characterized by a single wave
function) even at zero temperature.

Introducing the ratio $m=\frac{T}{T_{\text{eff}}}$ it follows $p_{\psi}\propto Z_{\psi}^{m}$,
leading to \begin{eqnarray*}
\widetilde{F} & = & \frac{\partial\left(mF\left(m\right)\right)}{\partial m}\\
S_{c} & = & \frac{m}{T_{\text{eff}}}\frac{\partial F\left(m\right)}{\partial m},\end{eqnarray*}
 where we introduced $F=\widetilde{F}-T_{\text{eff}}S_{c}$ with \[
F\left(m\right)=-\lim_{g\rightarrow0}\frac{T}{m}\log\int D\psi Z_{m}\left[\psi\right].\]
 It is now possible to integrate out the auxiliary variable $\psi$,
yielding an $m$-times replicated theory of the original variables
$\varphi$ with infinitesimal inter-replica coupling: \begin{equation}
\mathcal{S}=\sum_{\alpha=1}^{m}\mathcal{S}[\varphi^{\alpha}]-g\sum_{\alpha,\beta=1}^{m}\int d^{3}xd\tau d\tau^{\prime}\varphi_{\mathbf{x}}^{\alpha}\left(\tau\right)\varphi_{\mathbf{x}}^{\beta}\left(\tau^{\prime}\right),\label{repact}\end{equation}
 similar to a random field model with infinitesimal randomness $g$.
The major difference here is that in systems with a tendency towards
self-generated glassiness, the initial infinitesimal randomness $g$
will self consistently be replaced by an effective, interaction induced,
self-generated randomness. Specifically this will be the off diagonal
element of the self energy in replica space.

At this point it is useful to discuss similarities and differences
of the present approach if compared to the conventional replica approach
of systems with quenched disorder. First, on a technical level, the
replica index has to be analytically continued to $m=T/T_{\text{eff}}\leq1$
and not to zero. This reflects the fact that slow metastable configurations
do not equilibrate at the actual temperature, $T$, but at the effective
temperature $T_{\text{eff}}$. In other words, the system has an essentially
equal probability to evolve into states which are spread over a spectrum
with width $T_{\text{eff}}$, even as $T\rightarrow0$. If one applies
the present approach to a system with explicit quenched disorder,
where one can apply the conventional replica theory, and assumes replica
symmetry, it turns out that $m$ corresponds to the break point of
a solution with one step replica symmetry breaking of the conventional
replica approach and both techniques give identical results. Alternatively
one might also consider the model, Eq.\ref{qbraz}, in the limit of
an infinitesimal random field with width $g$ and break point of a
one step replica symmetry breaking $m$ solution. This implies that
the present approach captures the essence of glassy behavior in systems
with two very distinct typical time scales. This will become particularly
clear if we compare our results with the one obtained within the solution
of the dynamical Schwinger-Keldysh theory in the appendix. A more
physical relationship between the two replica approaches can be obtained
by realizing that the typical free energy of a frozen state, $\widetilde{F}$,
can also be written in the usual replica language via \[
\widetilde{F}=\lim_{n\rightarrow0}\frac{1}{n}\left(\overline{Z_{\psi}^{n}}-1\right)\]
 where the average is performed with respect to the distribution function
$p_{\psi}\propto Z_{\psi}^{m}$. The distribution function of the
self generated randomness is non-Gaussian. For example if $m\rightarrow1$,
$e^{-\mathcal{S}_{\psi}\left[\varphi\right]}$ can also be interpreted
as the generating functional of the distribution $p_{\psi}$.\cite{RMV01}
It is because this distribution is characterized by {}``colored noise''
that we find a self generated glassy state of the kind discussed here.
Finally, if one replaces $\left(\left[\nabla^{2}+q_{0}^{2}\right]\varphi\right)^{2}$
in Eq.\ref{qbraz} by the usual $\left(\nabla\varphi\right)^{2}$-term
(which is the $q_{0}\rightarrow0$ limit after appropriate rescaling
of $\varphi$, $c$, $u$. etc.) there is no glass as $g\rightarrow0$,
making evident that self-generated glassiness is ultimately caused
by the uniform frustration of the finite $q_{0}$ problem where modulated
configurations, $\varphi\left(\mathbf{x}\right)\varpropto\cos\left(\mathbf{q}_{0}\cdot\mathbf{x}\right)$,
with $\left|\mathbf{q}_{0}\right|=q_{0}$ and arbitrary direction
have low energy.

\begin{figure}
\includegraphics[%
  width=1.0\columnwidth,
  keepaspectratio]{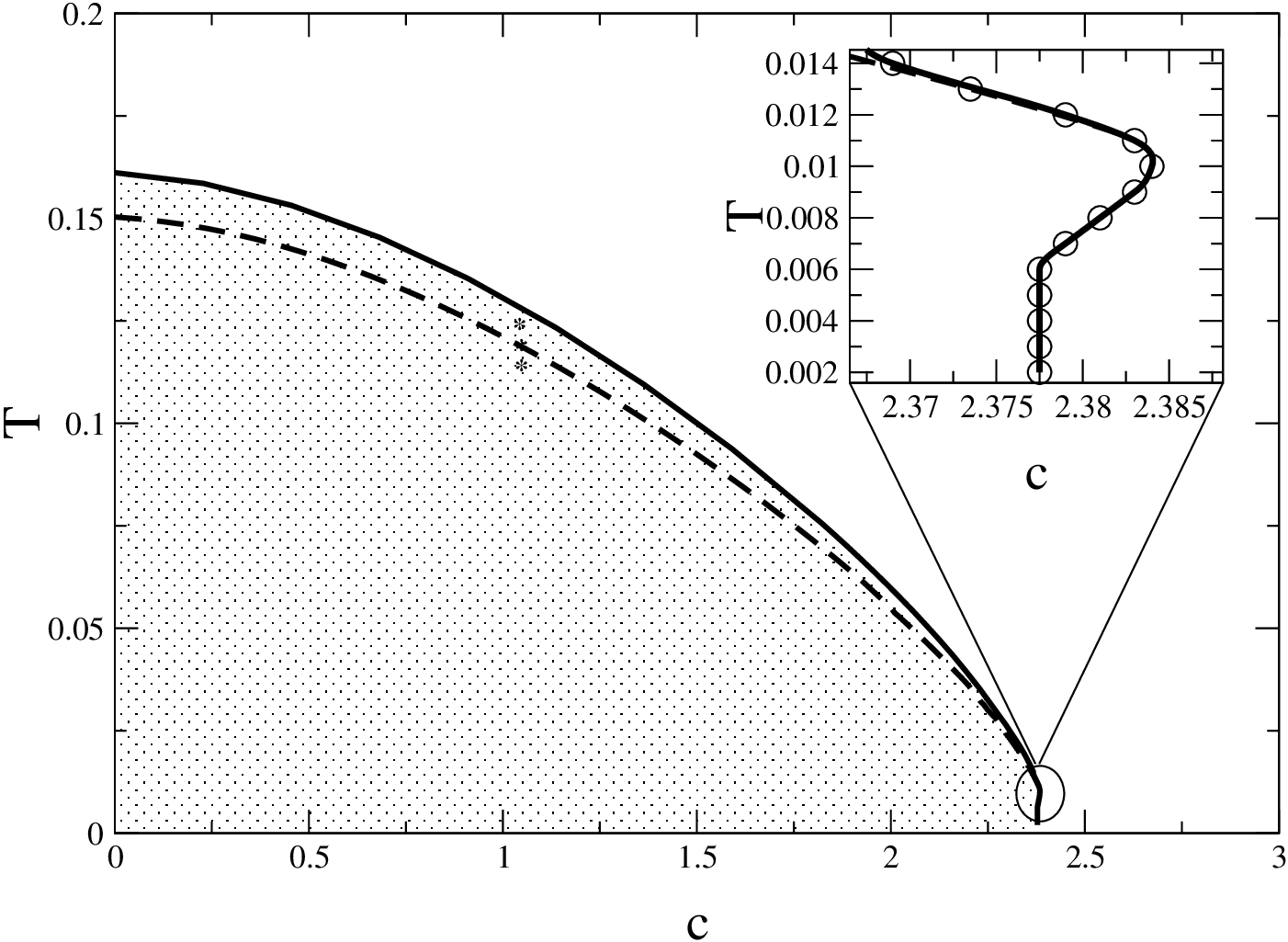}

\caption{\label{pdnew2} DMFT Phase diagram for the model Eq. 1 , within the
one-loop self consistent screening approximation. The solid line represents
$T_{\text{A}}$ whereas the dashed line represents $T_{\text{K}}$.
The inset shows the low temperature region of the phase diagram where
$T_{\text{A}}$ and $T_{\text{K }}$ have merged and the effective
temperature at the transition is larger than $T$ ($m(T_{\textrm{A}})<1$).
The line is simply a guide to the eye whereas the dots are the actual
numerical result. The numerical simulations in the paper where carried
out using the parameters $r_{0}=-6$, $q_{0}=0.3$ and $u=2\pi^{2}\left|r_{0}\right|$.
The stars refer to the $(T,c)$ points of which the results in Fig.4
are shown.}
\end{figure}

\section{Dynamical mean field theory}

Due to the mean field character of the theory, it is appropriate to
proceed by using the ideas of the DMFT for equilibrium many body systems\cite{dmftreview1,dmftreview2,dmftreview3}
and assuming that the self energy of our replicated field theory is
momentum independent. Physically, ignoring the momentum dependence
of the self energy might be justified by the fact that a glass transition
usually occurs in a situation of intermediate correlations, i.e. when
the correlation length of the liquid state is slightly larger but
comparable to the typical microscopic length scales in the Hamiltonian.\cite{SWW00}
The free energy of the system can be expressed in terms of the Matsubara
Green's function $G_{\mathbf{q}}^{\alpha\beta}\left(\omega_{n}\right)=\left\langle \varphi_{\mathbf{q}}^{a}(\omega_{n})\varphi_{-\mathbf{q}}^{b}(-\omega_{n})\right\rangle $
and the corresponding local (momentum independent) self energy $\Sigma^{\alpha\beta}\left(\omega_{n}\right)$
as \[
F\left(m\right)=\,\,\text{tr}\left(\Sigma G\right)-\text{tr}\log G+\Phi\left[G\right]\,\,,\]
 where the self energy is given by $\Sigma=\frac{\delta\Phi\left[G\right]}{\delta G}$,
and $\Phi\left[G\right]$ is diagrammatically well defined for a given
system\cite{KB61}.

The difference from the usual equilibrium DMFT approach\cite{dmftreview1,dmftreview2,dmftreview3}
is the occurrence of off diagonal elements in replica space, allowing
us to map the system onto a local problem with the same interaction
but dynamical {}``Weiss'' field matrix. This brings us to an effective
zero dimensional theory similar to the mode-coupling theory of classical
glasses.\cite{mc} We furthermore make the ansatz\cite{Mon95}\begin{equation}
G_{\mathbf{q}}^{\alpha\beta}\left(\omega_{n}\right)=\mathcal{G}_{\mathbf{q}}\left(\omega_{n}\right)\delta_{ab}+\mathcal{F}_{\mathbf{q}}\frac{\delta_{n,0}}{T},\label{GF}\end{equation}
 with static off diagonal elements. A similar ansatz for the self
energy leads to the following two Dyson equations for the diagonal
and off diagonal propagators: \begin{eqnarray}
\mathcal{G}_{\mathbf{q}}\left(\omega_{n}\right) & = & \left[\mathcal{G}_{0,\mathbf{q}}^{-1}\left(\omega_{n}\right)-\Sigma_{\mathcal{G}}\left(\omega_{n}\right)\right]^{-1},\nonumber \\
\mathcal{F}_{\mathbf{q}} & = & \Sigma_{\mathcal{F}}\mathcal{G}_{\mathbf{q}}\left(0\right)\left[\mathcal{G}_{\mathbf{q}}^{-1}\left(0\right)-m\Sigma_{\mathcal{F}}/T\right]^{-1}.\label{D}\end{eqnarray}
 Glassiness is associated with finite values of the Edwards-Anderson
order parameter, $\mathcal{F}_{\mathbf{q}}$, whereas for $\mathcal{F}_{\mathbf{q}}=0$
we recover the traditional theory of quantum liquids. The structure
of the Dyson equation already gives us crucial informations on the
nature of the glass transition. From Eq.\ref{D} follows immediately
that, contrary to the classical case where glassiness can occur with
$m=1$\cite{SW00,SWW00}, in the quantum limit ($T\rightarrow0$)
the only way $\mathcal{F}_{\mathbf{q}}$ can be non-zero is to have
$m\rightarrow0$ such that $m/T=1/T_{\text{eff}}\neq0$. This imposes
a constraint on the replica symmetry breaking structure in the quantum
limit of the replica approach developed in Ref.\onlinecite{Mon95}.
Also, it is clear that $\Sigma_{\mathcal{F}}$ defines a new length
scale of the problem that is associated with the glass transition.\cite{SWW00}
In the classical glass transition the relation $m=1$ is satisfied
at $T_{A}$. Then the two Dyson equations can be decoupled into an
equilibrium, diagonal (in replica space) part, and a non-equilibrium,
off-diagonal part. This allows us to interpret $\Sigma_{\mathcal{G}}$
as related with the equilibrium correlation length and to associate
$\Sigma_{\mathcal{F}}$ with the Lindemann length, associated with
the typical length scale of wandering of defects of the equilibrium
structure, see Ref.\cite{SWW00} for details. However, it follows
from the structure of the self-energies that, in the quantum limit,
where $m<1$ at $T_{A}$, the two Dyson equations cannot be decoupled
anymore. In this case, the two self-energies will combined define
a correlation length and a Lindemann length which are not independent,
but rather closely intertwined.

We solved the impurity problem within a self consistent large-$N$
approach, i.e. we generalize the scalar field $\varphi$ to an $N$-component
vector and consider the limit of large-$N$ including first $1/N$-corrections.
This approach was used earlier to investigate self generated glassiness
in the classical limit\cite{SW00,SWW00}. In this limit it is also
possible to solve the DMFT-problem exactly\cite{WSKW02}, demonstrating
that glassiness found in the approximate large-$N$ limit is very
similar to the exact, finite $N$ theory and thereby supporting the
applicability of the large $N$-expansion.

Within the self consistent large-$N$ approach, the diagonal and off
diagonal element of the self energy are given as \begin{eqnarray}
\Sigma_{\mathcal{G}}\left(\omega_{n}\right) & = & \Sigma_{\mathcal{G}}^{H}-\frac{2}{N}\left(\mathcal{F}D_{\mathcal{G}}\left(\omega_{n}\right)+D_{\mathcal{F}}\mathcal{G}\left(\omega_{n}\right)\right)\nonumber \\
 &  & +T\sum_{m}D_{\mathcal{G}}\left(\omega_{n}+\omega_{m}\right)\mathcal{G}\left(\omega_{m}\right)\nonumber \\
\Sigma_{\mathcal{F}} & = & -\frac{2}{N}D_{\mathcal{F}}\mathcal{F},\label{scsa1}\end{eqnarray}
 with Hartree contribution \[
\Sigma_{\mathcal{G}}^{H}=-uT\sum_{m}\mathcal{G}\left(\omega_{m}\right)-u\mathcal{F}\]
 as well as \begin{eqnarray}
D_{\mathcal{G}}\left(\omega_{n}\right) & = & \frac{1}{u^{-1}+\Pi_{\mathcal{G}}\left(\omega_{n}\right)+2\gamma\mathcal{FG}\left(\omega_{n}\right)}\nonumber \\
D_{\mathcal{F}} & = & \frac{-\gamma\mathcal{F}^{2}D_{\mathcal{G}}^{2}\left(0\right)}{1+\gamma\mathcal{F}^{2}D_{\mathcal{G}}\left(0\right)/T_{\text{eff}}}\label{scsa_2}\end{eqnarray}
 and bubble diagram $\Pi_{\mathcal{G}}\left(\omega_{n}\right)=\gamma T\sum_{m}\mathcal{G}\left(\omega_{n}+\omega_{m}\right)\mathcal{G}\left(\omega_{m}\right)$.
Here $\mathcal{G}\left(\omega_{m}\right)=\int\frac{d^{3}q}{\left(2\pi\right)^{3}}\mathcal{G}_{\mathbf{q}}\left(\omega_{n}\right)$
and $\mathcal{F}=\int\frac{d^{3}q}{\left(2\pi\right)^{3}}\mathcal{F}_{\mathbf{q}}$
are the momentum averaged propagators. The DMFT is usually formulated
on a lattice and it is possible to chose the same dimension (e.g.
inverse energy) for the momentum dependent and momentum averaged propagator,
by assuming the lattice spacing equal unity. In a continuum theory
the role of the lattice spacing is played by the inverse upper cut
off of the momentum integration $\Lambda$, which enters our theory
in Eqs.\ref{scsa_2} through the constant $\gamma=\Lambda^{-3}$.
In the case of the Hamiltonian, Eq.\ref{qbraz}, all momentum integrals
are convergent as $\Lambda\rightarrow\infty$ and the scale which
replaces the cut off is $q_{0}$, leading to $\gamma=q_{0}^{-3}$.
We solved this set of self consistent equations numerically. Before
we present the results we must discuss the stability of the ansatz,
Eq.\ref{GF}.

The local stability of the replica symmetric ansatz, Eq.\ref{GF}
is determined by the lowest eigenvalue of the Hessian matrix $\frac{\delta^{2}F}{\delta G^{\alpha\beta}\delta G^{\gamma\delta}}$.
Proceeding along the lines of Ref.\onlinecite{dAT78} and diagonalization
over the replica indices leads to the following matrix in momentum
space \begin{equation}
\mathcal{M}_{\mathbf{q},\mathbf{q}^{\prime}}=\delta\left(\mathbf{q}-\mathbf{q}^{\prime}\right)\mathcal{G}_{\mathbf{q}}^{-2}\left(0\right)+\mathcal{C}\,,\label{repliconseq}\end{equation}
 where \begin{equation}
\mathcal{C}=\frac{\delta^{2}\Phi}{\delta G^{\alpha\beta}\delta G^{\alpha\beta}}-2\frac{\delta^{2}\Phi}{\delta G^{\alpha\beta}\delta G^{\alpha\delta}}+\frac{\delta^{2}\Phi}{\delta G^{\alpha\beta}\delta G^{\gamma\delta}},\label{deriative}\end{equation}
 with distinct $\alpha,\beta,\delta$ and $\delta$. Diagrammatically,
$\mathcal{C}$ is a sum of diagrams with four external legs with at
least two distinct replica indices. Thus, as $g\rightarrow0$ the
constant $\mathcal{C}$ vanishes if $\mathcal{F}_{\mathbf{q}}\rightarrow0$,
an observation which will be relevant in our discussion of the nature
of the zero temperature glass transition as a function of $c$. We
find for the lowest eigenvalue, $\lambda$, of $\mathcal{M}_{\mathbf{q},\mathbf{q}^{\prime}}$:
\begin{equation}
\mathcal{C}^{-1}=\int\frac{d^{3}q}{\left(2\pi\right)^{3}}\left(\mathcal{G}_{\mathbf{q}}^{-2}\left(0\right)+\lambda\right)^{-1}.\label{stability}\end{equation}
 If $\lambda>0$ the mean-field solution is stable, whereas it is
marginal for $\lambda=0$. The fairly simple result, Eq.\ref{stability},
for the stability of the replica structure is a consequence of the
momentum independence of the self energy within DMFT which guarantees
that $\Phi\left[G\right]$ only depends on the momentum averaged propagators.
Otherwise $\mathcal{C}$ would depend on momentum, making the analysis
of the eigenvalues of $\mathcal{M}_{\mathbf{q},\mathbf{q}^{\prime}}$
in Eq.\ref{repliconseq} much more complicated. Thus, the investigation
of our problem within dynamical mean field theory is not only convenient
to obtain solutions for $\mathcal{G}_{\mathbf{q}}\left(\omega_{n}\right)$
and $\mathcal{F}_{\mathbf{q}}$, it is also crucial to make progress
in the analysis of the stability of this solution. We expect that
it will be impossible to find a stable replica symmetric solution,
Eq.\ref{GF}, once one goes beyond DMFT.

In real physical systems the slow degrees of freedom relax on a finite
time scale $\tau_{\alpha}$ and the effective temperature $T_{\text{eff}}=T/m$
depends not only on the external parameters like $T$ and pressure,
but also on the cooling rate, or equivalently on the time, $t_{w}$,
elapsed after quenching and thus on the events which, for a finite
range system, can take the system to different states of the spectrum
characterized by $T_{\text{eff}}$. Accounting for these effects goes
beyond a mean field treatment. However, for $t_{w}\ll\tau_{\alpha}$
mean field theory should apply. In fact, since within the mean field
approach the time $\tau_{\alpha}$ is infinite, this is intrinsically
the regime we are constrained to within the mean field treatment.
Then, the most important configurations of the order parameter are
those which allow the system to explore the maximum number of ergodic
regions. The best way to achieve this, without being unstable, is
through interconnecting saddle points.\cite{AC01} This leads to the
marginal stability condition, $\lambda=0$, which we use to determine
the effective temperature $T_{\text{eff}}^{0}$, which corresponds
to the effective temperature right after a fast quench into the glassy
state, i.e, $T_{\text{eff}}^{0}=T_{\text{eff}}(t_{w}\ll\tau_{\alpha})$:
\begin{equation}
\mathcal{C}\left(T_{\text{eff}}^{0}\right)^{-1}=\int\frac{d^{3}q}{\left(2\pi\right)^{3}}\mathcal{G}_{\mathbf{q}}^{2}\left(0,T_{\text{eff}}^{0}\right).\label{Teff}\end{equation}
 Within the large-$N$ approximation used to determine the propagator
we can also evaluate the constant $\mathcal{C}$ of Eq.\ref{deriative},
leading to \begin{equation}
\mathcal{C}\left(T_{\text{eff}}^{0}\right)=\frac{2}{N}\left[2D_{\mathcal{G}}^{2}\left(0\right)\Pi_{\mathcal{F}}-\mathcal{D}_{\mathcal{F}}\right]\label{Teff2}\end{equation}
 which is an implicit equation for $T_{\text{eff}}^{0}$. Note that,
as we argued before, $\mathcal{C}$ vanishes for $\mathcal{F}\rightarrow0$.
Moreover, since the integral on the right hand side of Eq.\ref{Teff}
is bounded from above, $\mathcal{C}$ and therefore $\mathcal{F}_{\mathbf{q}}$
must vanish discontinuously at the transition even for $T=0$. As
explained above, from the structure of the Dyson equation in replica
space it immediately follows that for $c$ larger than some value
$c_{A}$, $T_{\text{eff}}^{0}/T$ must also jump discontinuously from
$T_{\text{eff}}^{0}/T>1$ in the glass state to $T_{\text{eff}}^{0}/T=1$
in the quantum liquid state.

\begin{figure}
\psfrag{F}{$\mathcal{F}$}
\includegraphics[%
  width=1.0\columnwidth,
  keepaspectratio]{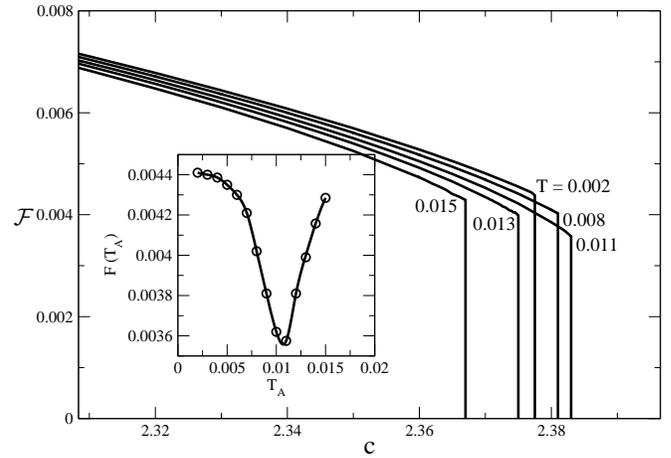}

\caption{\label{Fofc} Low temperature behavior of the Edwards-Anderson order
parameter as a function of the intensity of quantum fluctuations.
The temperatures are in the region displayed in the inset of Fig.
\ref{pdnew2}. Inset: dependence of the order parameter at the transition
temperature $T_{\text{A}}$.}
\end{figure}

Together with Eq.\ref{Teff} we have a closed set of equations which
describe the quantum glass within mean field theory. We have solved
this set of equations for the model, Eq.\ref{qbraz}, and indeed found
that there is a glassy state below a critical value for $c$. Most
interestingly, for increasing $c$ (i.e.increasing quantum fluctuations)
the rapidly quenched quantum glass, which at some point becomes unstable,
undergoes a discontinuous reorganization of the density matrix upon
entering the quantum liquid state. It is crucial to solve this many-body
problem within some conserving approximation, i.e., based upon a given
functional $\Phi\left(G\right)$, which must be simultaneously used
to determine $\Sigma$ and $\mathcal{C}$. The set of equations was
solved numerically. The Matsubara frequency convolutions were calculated
on the imaginary time axis via Fast Fourier Transform algorithm using
$2^{15}$ frequencies. This accuracy is needed mostly to be able to
find solutions of Eqs.\ref{Teff} and \ref{Teff2} .

\begin{figure}
\includegraphics[%
  width=1.0\columnwidth,
  keepaspectratio]{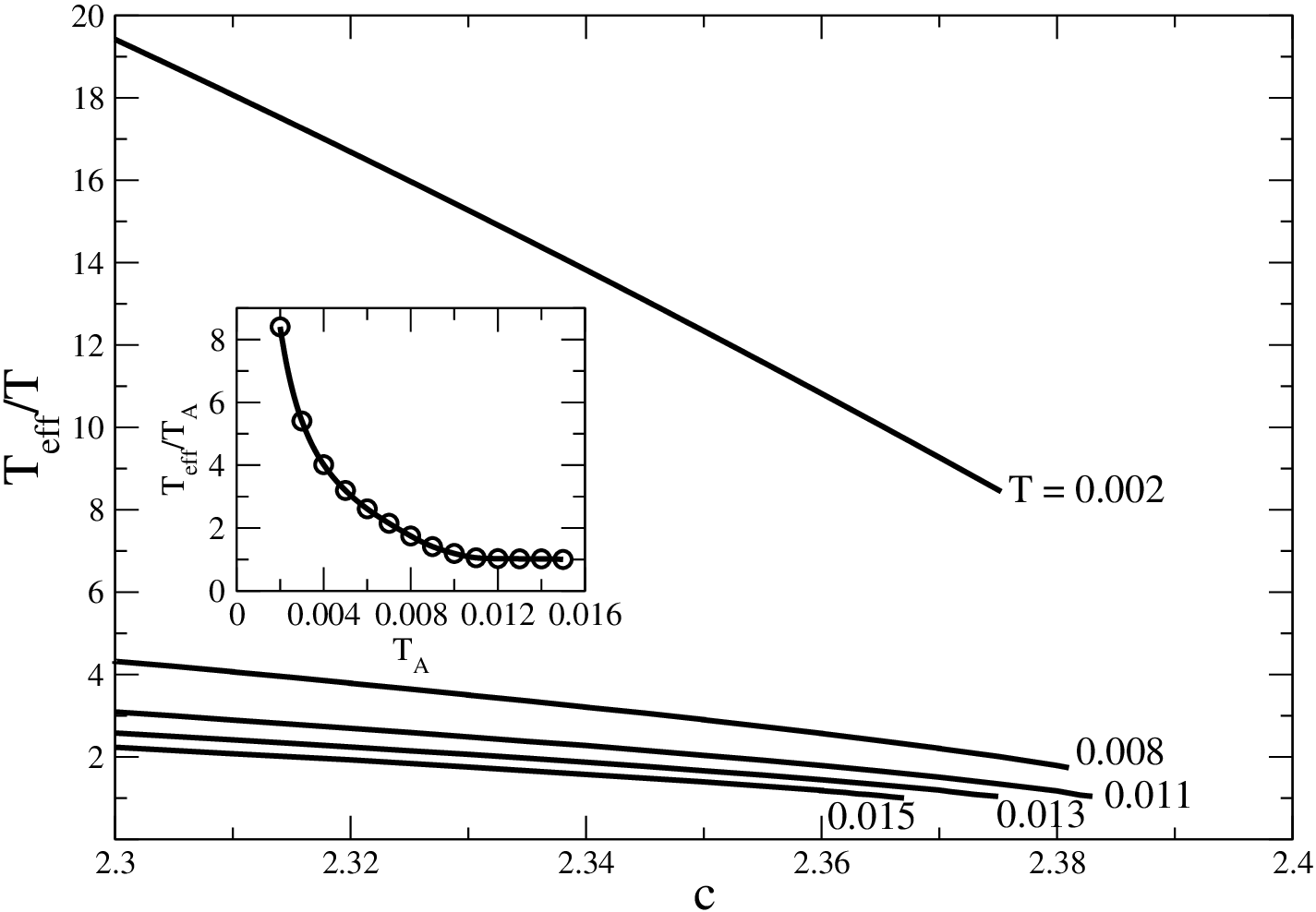}

\caption{\label{mofc} Dependence of the effective temperature as function
of $c$ for different temperatures. Inset: dependence of $T_{\text{eff}}^{0}/T_{\text{A}}$
on $T_{\text{A}}$ in the low temperature regime displayed on the
inset of Fig. \ref{pdnew2}.}
\end{figure}

The transition line between the liquid and glassy states in the $(c,T)$
space is presented on Fig \ref{pdnew2}. Note that the low temperature
(quantum regime) behavior is qualitatively different from the classical
stripe glass. In the quantum limit, $T_{\text{A}}$ and $T_{\text{K}}$
merge and the effective temperature at the transition $T_{\text{eff }}^{0}\left(T_{\text{A}}\right)$
is always larger than $T_{\text{A}}$, i.e., $m<1$. Due to the reentrant
character of the transition the quantum glass can also be reached
by heating up the system. By generalizing Brazovskii's theory of the
fluctuation induced first order transition to the quantum case, we
found a similar reentrance behavior for this equilibrium transition,
suggesting that this peculiar shape of the phase border is determined
by the increasing relevance of fluctuations with wave vector $\mathbf{q}=\left|q_{0}\right|$
as one crosses over from a quantum to a classical regime (see corresponding
remarks made in sec.II). Another way the quantum glass can be reached
is of course via a {}``$c$ quench''. Note also that the reentrant
behavior we find within our approach happens at the point where numerically,
at the same time, $T_{\text{A}}-T_{\text{K}}$ vanishes (to within
numerical precision), $m\left(T_{\text{A}}\right)$ starts falling
with a larger derivative (see the inset of Fig.\ref{mofc}), and $\mathcal{F}\left(T_{\text{A}}\right)$
reaches a minimum (see inset of Fig. \ref{Fofc}) before plateauing
as $T\rightarrow0$. This suggests that the reentrance behavior and
the change of character of the transition are closely related. In
Fig. 3 we show the dependence of the replica symmetry index $m=T/T_{\text{eff}}^{0}$
as function of $c$ for different temperatures.

In the classical limit, $c\rightarrow0$, $\mathcal{F}_{\mathbf{q}}$
changes discontinuously at $T_{A}$, whereas $T_{\text{eff}}$ changes
continuously $(m=1)$, i.e. the relevant metastable states - within
which a classical glassy system gets trapped into - connect gradually
to the relevant states which contribute to the liquid state partition
function. A similar behavior in the quantum limit would imply that
$T_{\text{eff}}$ and therefore $\mathcal{F}_{\mathbf{q}}$ and $\mathcal{C}$
vanish continuously at the quantum glass transition. However, as discussed
above, this is not possible if one uses Eq.\ref{Teff} to determine
the effective temperature. Thus, $T_{\text{eff}}$ changes discontinuously
and one might expect a nucleation of liquid droplets within the unstable
glass state to be important excitations which cause a quantum-melting
of the glass. Even though one can formally introduce, along the lines
of Ref.\onlinecite{N98}, a latent heat $\delta Q=T_{\text{eff}}S_{c}$
at this transition, we do not know of a scenario which, within mean
field theory, allows this energy to be realized within the laboratory.
The appearance of a first order like transition as one enters the
quantum regime, however without reentrance behavior, was first pointed
out in Ref.\onlinecite{Cugl} in a related case of spin glasses with
quenched disorder.

Finally, we make contact between our theory and the Schwinger-Keldysh
approach used in Ref.\onlinecite{CL99}, which gives a set of coupled
equations for the symmetrized correlation function $C_{\mathbf{x},\mathbf{x}^{\prime}}\left(\tau,t_{\text{w}}\right)=\frac{1}{2}\left\langle \left[\varphi_{\mathbf{x}}\left(t_{\text{w}}+\tau\right),\varphi_{\mathbf{x}^{\prime}}\left(t_{\text{w}}\right)\right]_{+}\right\rangle $
and the retarded response function $G_{\mathbf{x},\mathbf{x}^{\prime}}^{r}\left(\tau,t_{\text{w}}\right)=-i\theta\left(\tau\right)\left\langle \left[\varphi_{\mathbf{x}}\left(t_{\text{w}}+\tau\right),\varphi_{\mathbf{x}^{\prime}}\left(t_{\text{w}}\right)\right]_{-}\right\rangle $,
where $\left[\,,\,\right]_{\pm}\equiv AB\pm BA$ . In the classical
limit $C_{\mathbf{x},\mathbf{x}^{\prime}}\left(\tau,t_{\text{w}}\right)=\left\langle \varphi_{\mathbf{x}}\left(t_{\text{w}}+\tau\right)\varphi_{\mathbf{x}^{\prime}}\left(t_{\text{w}}\right)\right\rangle $as
usual. If $t_{\text{w}}$, is comparable to $\tau$ the dynamics is
complex and depends on the nature of the initial state (aging regime).
On the other hand, for $t_{\text{w}}$ large compared to $\tau$,
$C\left(\tau,t_{\text{w}}\right)$ and $G^{r}\left(\tau,t_{\text{w}}\right)$
are expected to dependent only on $\tau$ (stationary regime). Correspondingly,
one can decompose the correlation function into aging and stationary
contributions \begin{equation}
C\left(\tau,t_{\text{w}}\right)=C_{\text{AG}}\left(\tau,t_{\text{w}}\right)+C_{\text{ST}}\left(\tau\right)\,,\label{splitagst}\end{equation}
 and similarly for $G^{r}\left(\tau,t_{\text{w}}\right)$. Cugliandolo
and Kurchan\cite{CK93} showed within the mean field theory of classical
spin glasses that one cannot decouple the stationary dynamics from
the aging regime. Instead, the system establishes a {}``weak long
term memory'' and one has to solve for the entire time dependence.
An elegant way to encode the aging dynamics is a generalized fluctuation
dissipation theorem (FDT) $G_{\text{AG}}^{r}\left(\tau,t_{\text{w}}\right)=-T_{\text{eff}}^{-1}\frac{\partial}{\partial t_{\text{w}}}C_{\text{AG}}\left(\tau,t_{\text{w}}\right)$,
with effective $T_{\text{eff}}$. This approach was generalized to
the quantum case in Ref.\onlinecite{CL99}.

As shown in the appendix, we find a complete equivalence of the Schwinger-Keldysh
theory of Ref.\onlinecite {CL99}, if applied to the quantized Brazovskii
model, with our replica approach if we identify (after analytical
continuation to real time): $\mathcal{G}\left(t\right)=G_{\text{ST}}^{r}\left(t\right)$and
$\mathcal{F}=\lim_{t_{\text{w}}\rightarrow\infty}\lim_{t\rightarrow\infty}C_{\text{AG}}\left(t_{\text{w}},t\right)$.
$T_{\text{eff}}^{0}$, which follows from marginality, Eq.\ref{Teff},
is identical to the one of the generalized FDT, when we assume $t_{\text{w}}\ll\tau_{\alpha}$.\cite{BC01}

\section{Aspects beyond the strict mean field theory}

In this section we discuss several physically significant conclusions
one can draw from our theory which go beyond the strict mean field
limit. In particular we will discuss several aspects related to dynamical
heterogeneity in glasses.

Going beyond mean field theory, for $t_{w}\gg\tau_{\alpha}$, the
marginal stability cannot be sustained, since the system can save
free-energy via droplet formation,\cite{KTW89} which drives the system
towards equilibrium. As discussed in Ref.\onlinecite{KTW89}, the
free-energy gain is due to a gain in configurational entropy which
is inhibited within mean field theory due to infinitely large barriers,
but possible in finite subsystems where transitions between distinct
metastable states becomes allowed. Thus, the glass might be considered
as consisting of a mosaic pattern built of distinct mean field metastable
states. The size of the various droplets forming the mosaic is determined
by a balance of the entropic driving force (proportional to the volume
of the drop) and the surface tension\cite{KTW89}.

One way to account phenomenologically for such behavior within our
theory is to assume that $T_{\text{eff}}$ becomes a time dependent
quantity\cite{N98} and the exploration of phase space allows the
system to {}``cool down'' its frozen degrees of freedom by realizing
configurational entropy. One would naturally expect $T_{\text{eff}}(t_{w})$
to decrease towards a value $T_{\text{eff}}^{\infty}$ until either
$\lim_{N\rightarrow\infty}S_{c}(T_{\text{eff}}^{\infty})/N=0$ or
$T_{\text{eff}}^{\infty}=T$. In the former case there is no extensive
entropic driving force anymore which favors the exploration of phase
space, whereas in the latter case the system has reached equilibrium
but with, in general, finite remaining configurational entropy. These
two regimes are separated by the Kauzmann temperature where $\lim_{N\rightarrow\infty}S_{c}(T_{\text{eff}}^{\infty})/N=0$
at $T_{\text{eff}}^{\infty}=T$ simultaneously\cite{KTW89}.We cannot,
of course, calculate the explicit time dependence of $T_{\text{eff}}(t_{w})$
here. However, we can parametrically study $S_{c}(T_{\text{eff}})$
versus $T_{\text{eff}}$, i.e. keep the replica value $m$ an open
parameter of the theory and analyze whether the trends for the variation
of $S_{c}(T_{\text{eff}})$ at fixed temperature $T$ are sensible.
Since we determined $m$ previously by the marginality condition,
an effective temperature $T_{\text{eff}}<T_{\text{eff}}^{0}$ implies
that the replicon eigenvalue is different from zero. We found numerically
that $\lambda<0$, reflecting an instability of our solution likely
related to some kind of dynamical heterogeneity. We argue that this
heterogeneity is different for temperatures close to $T_{A}$ and
$T_{K}$.%
\begin{figure}
\includegraphics[%
  width=1.0\columnwidth,
  keepaspectratio]{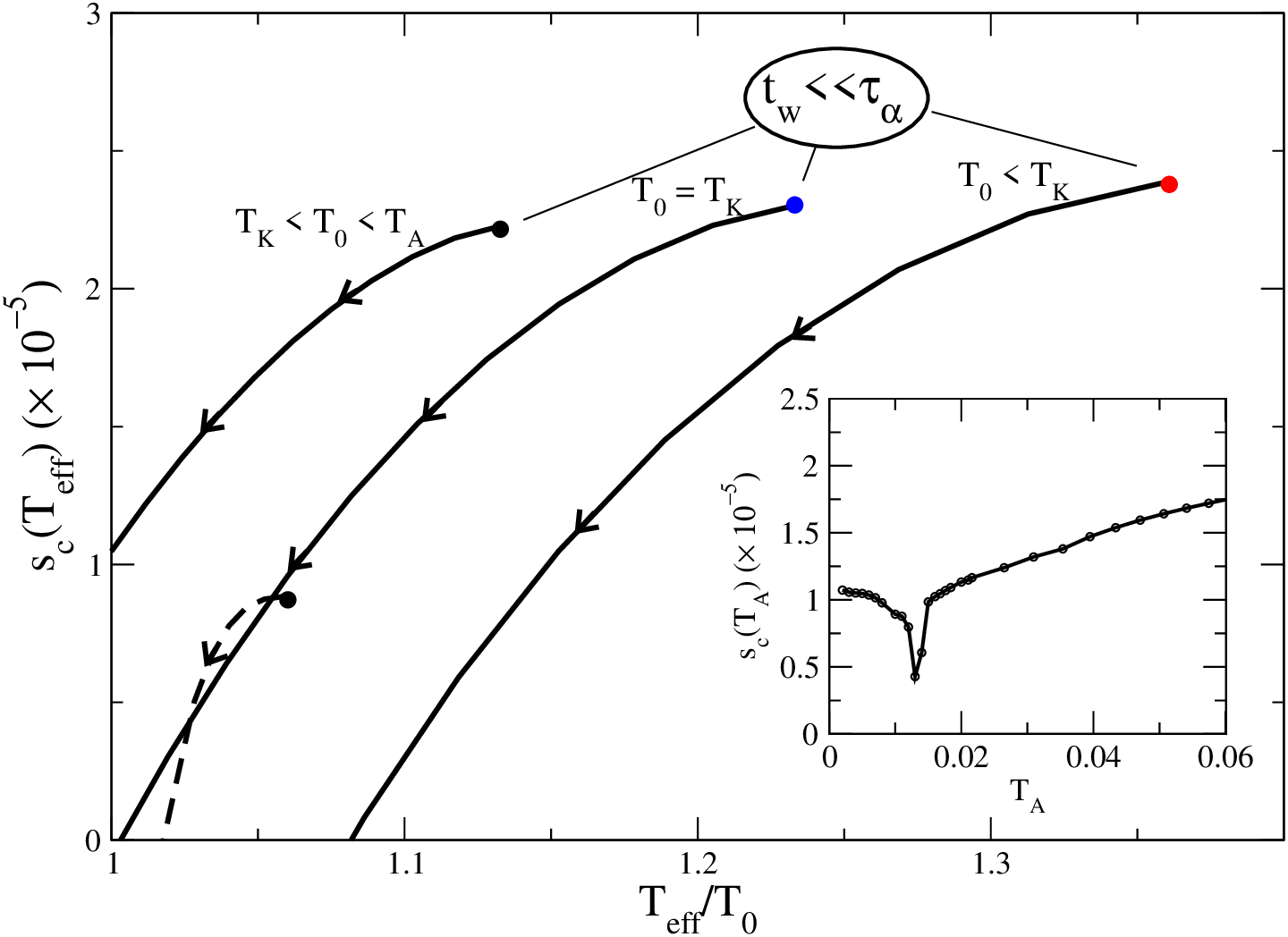}

\caption{\label{ScofTa} Parametric plot of the relaxation of $T_{\text{eff}}$
and $S_{c}$ for varying $t_{w}$ after quenches to the 3 points indicated
by stars in the phase diagram Fig.\ref{pdnew2}. In addition $S_{c}$
is shown for one point right at the quantum glass transition with
$T_{A}=0.01>T$. The direction of evolution of the waiting time $t_{w}$
is indicated by the arrows. The system can relax from a state with
marginal stability with $T_{\text{eff}}^{0}$ to 3 different final
states as explained in the text. Inset: configurational entropy at
the transition point as function of $T_{\text{A}}$.}
\end{figure}

We first discuss the behavior below but close to $T_{A}$ in the classical
regime where marginality gives a continuous change of $m$, i.e. $T_{\text{eff}}^{\infty}\left(T_{A}\right)=T$.
In this regime the system is close to equilibrium and it should be
sensible to consider small fluctuations around the mean field solution.
Such small fluctuations are then dominated by the eigenvectors of
the replicon problem which correspond to the lowest eigenvalue. From
Eq.\ref{repliconseq} we can easily determine the momentum dependence
of this eigenvector, which can be interpreted as the fluctuating mean
square of the long time correlation function $\mathcal{F}_{\mathbf{q}}$
away from its mean field value. It is given by \begin{equation}
\Psi_{\mathbf{q}}=\Psi_{0}\frac{q_{0}^{4}}{\mathcal{G}_{\mathbf{q}}^{-2}-\lambda}\label{eigenval}\end{equation}
 where, $\Psi_{0}$ is a normalization constant. At marginality, these
modes in correlation function space are massless and thus easy to
be excited. The typical length scale of these correlation function
fluctuations are determined by $\mathcal{G}_{\mathbf{q=q}_{0}}^{-2}$,
i.e. are confined to a length scale determined by $\Sigma_{\mathcal{G}}^{-1/2}$.
Since this length is not the actual correlation length, but rather
determined by the shorter Lindemann length discussed on Ref.\onlinecite{SWW00},
the wandering of defects seem to set the scale on which the dynamics
close to $T_{A}$ evolves. Thus correlations over the Lindemann length
are fluctuating in space. Even though the fluctuating object is characterized
by a rather short scale, its fluctuations are, of course, correlated
over larger distances, as characterized by the nonlinear susceptibility
$\chi^{\alpha\beta\gamma\delta}=\left\langle \varphi_{a}\left(x\right)\varphi_{\beta}\left(x\right)\varphi_{\gamma}\left(x\right)\varphi_{\delta}\left(x\right)\right\rangle $.
$\chi^{\alpha\beta\gamma\delta}$ is the inverse of the Hessian and
thus diverges if $\lambda\rightarrow0$. Thus we conclude that close
to $T_{A}$ this heterogeneity is driven by the correlation function
fluctuations of shape $\Psi_{\mathbf{q}}$ of Eq.\ref{eigenval}.
We believe that this {}``Goldstone'' - type heterogeneity is very
similar in character to the recent interesting approach to heterogeneity
resulting from the assumption of a local time reparametrization invariance\cite{Castillo02}.
This is supported by the close relation between the reparametrization
invariance and marginality as shown in Ref.\onlinecite{CDD}. In addition,
the approach of Ref.\onlinecite{Castillo02} considers fluctuations
relative to the mean field solution with stiffness proportional to
the (mean field) Edwards-Anderson parameter. Thus, small fluctuations
caused by the marginality of the mean field solution are considered,
similar to the ones given in Eq.\ref{eigenval} Further away from
$T_{A}$ such a linearized theory is likely to break down because
additional eigenvectors, not related to the marginal eigenvalue, become
relevant and non-Gaussian fluctuations come into play. This should
always be the case where $m$ is not close to unity, i.e. in the classical
regime for temperatures below $T_{A}$ and everywhere in the quantum
regime. One might expect droplet-physics to become important then\cite{KTW89}.

In both cases it is useful to analyze the evolution of the spectrum
of states of this formally unstable theory, particularly if the replica
structure of the theory remains unchanged. As shown in Fig.1, we found
numerically for the model, Eq.\ref{qbraz}, that for $t_{w}\gg\tau_{\alpha}$,
three different possible final situations, result, depending on the
relation between the bath temperature $T$ and the Kauzmann temperature
$T_{K}$: (1) $T_{A}>T>T_{K}$, $T_{\text{eff}}$ will relax until
it reaches $T$, but an excess entropy will remain. (2) $T=T_{K}$,
$T_{\text{eff}}$ will relax until it reaches $T$, in a state with
zero configurational entropy. (3) $T<T_{K}$, $T_{\text{eff}}$ will
relax until all the excess entropy vanishes, but the system remains
in a non-equilibrium state with $T_{\text{eff}}^{\infty}>T$. In this
case there are still many (even though less than exponentially many)
states, distributed according to a Boltzmann function with an effective
temperature, $T_{\text{eff}}^{\infty}$. Their energies, $\tilde{f}_{\psi}$,
differ by non-extensive amounts in a range of order $T_{\text{eff}}^{\infty}<T_{\text{eff}}^{0}$.
At a critical value of the quantum parameter slightly below $c_{A}$
we find that $T_{A}$ and $T_{K}$ merge and the nature of the glass
transition changes. The system is either in a quantum fluid or in
a non-equilibrium frozen state and the identification of the glass
transition using equilibrium techniques alone becomes impossible.
Note however that in the quantum glass regime, even at temperatures
$T$ arbitrarily close to $T_{\text{A}}$, for $t_{w}\gg\tau_{\alpha}$
we always obtain $\lim_{N\rightarrow\infty}S_{c}(T_{\text{eff}}^{\infty})/N=0$
for $T_{\text{eff}}^{\infty}>T$ (see dashed curve on Fig. \ref{ScofTa}).

\section{Conclusion}

In summary, we have presented an approach to self generated quantum
glasses which enables the counting of competing ground state energies
(quasi-classically of long lived metastable states) in interacting
quantum systems. Technically very similar in form to the traditional
quantum many body theory of equilibrium systems, it allows one to
investigate whether a given system exhibits self generated glassiness
as a consequence of the frustrating interactions. Slow degrees of
freedom are assumed to behave classically and are shown to equilibrate
to an effective temperature which is nonzero even as $T\rightarrow0$
and which characterizes the width and rigidity of the energy landscape
of the competing states of the system. Applied to the specific model,
Eq.\ref{qbraz}, we do find a glass below a critical value for the
quantum fluctuations. Using a marginality criterion to determine $T_{\text{eff}}$,
we can generally show that quantum glass transitions are bound to
be discontinuous transitions from pure to mixed quantum states. This
leads to the interesting question of how quantum melting of the non-equilibrium
quenched states can occur via nucleation of the corresponding quantum
liquid state. Even going beyond mean field theory by assuming a time
dependent effective temperature, we find that $T_{\text{eff}}$ saturates
(at least for extremely long times) at a value $T_{\text{eff}}^{\infty}>T$.
Finally we made connection to the dynamical approach for non-equilibrium
quantum many body systems of Ref.\onlinecite{CL99}, which shows that
our theory properly takes into account the effects of aging and long
term memory. We believe that the comparable simplicity of our approach
allows it to apply our technique to a wide range of interesting problems
in strongly interacting quantum systems.

This research was supported by an award from Research Corporation
(J.S.), the Institute for Complex Adaptive Matter, the Ames Laboratory,
operated for the U.S. Department of Energy by Iowa State University
under Contract No. W-7405-Eng-82 (H.W.Jr. and J. S.), and the National
Science Foundation grant CHE-9530680 (P. G. W.). H.W.Jr. acknowledges
support from FAPESP project No. 02/01229-7

\appendix

\section{Schwinger-Keldysh Formalism}

In this section we apply the Schwinger and Keldysh close-time path
Green function formalism applied to quantum glasses without quenched
disorder. We follow closely Cugliandolo and Lozano\cite{CL99}, who
applied the technique to spin glasses with quenched disorder. In this
formalism response and correlation functions are treated as independent
objects, coupled through a set of equations called Schwinger-Dyson
equations. A perturbation theory scheme can be set up by considering
the generalized matrix Green function\begin{equation}
\widetilde{G}_{\mathbf{x},\mathbf{x}^{\prime}}\left(t,t^{\prime}\right)=\left(\begin{array}{cc}
\,\,0 & G_{\mathbf{x},\mathbf{x}^{\prime}}^{a}\left(t,t^{\prime}\right)\\
G_{\mathbf{x},\mathbf{x}^{\prime}}^{r}\left(t,t^{\prime}\right) & C_{\mathbf{x},\mathbf{x}^{\prime}}\left(t,t^{\prime}\right)\end{array}\right)\label{retmat}\end{equation}
 and self-energy\[
\widetilde{\Sigma}_{\mathbf{x},\mathbf{x}^{\prime}}\left(t,t^{\prime}\right)=\left(\begin{array}{cc}
\,\,0 & \Sigma_{\mathbf{x},\mathbf{x}^{\prime}}^{a}\left(t,t^{\prime}\right)\\
\Sigma_{\mathbf{x},\mathbf{x}^{\prime}}^{r}\left(t,t^{\prime}\right) & \Sigma_{\mathbf{x},\mathbf{x}^{\prime}}\left(t,t^{\prime}\right)\end{array}\right).\]
 where \begin{eqnarray*}
C_{\mathbf{x},\mathbf{x}^{\prime}}\left(t,t^{\prime}\right) & = & \frac{1}{2}\left\langle \left[\varphi_{\mathbf{x}}\left(t_{\text{w}}+\tau\right),\varphi_{\mathbf{x}^{\prime}}\left(t_{\text{w}}\right)\right]_{+}\right\rangle \\
G_{\mathbf{x},\mathbf{x}^{\prime}}^{r(a)}\left(t,t^{\prime}\right) & = & \mp i\theta\left(\pm\left(t-t^{\prime}\right)\right)\left\langle \left[\varphi_{\mathbf{x}}\left(t\right),\varphi_{\mathbf{x}'}\left(t^{\prime}\right)\right]\right\rangle \end{eqnarray*}

In analogy to conventional perturbation theory, the components of
$\widetilde{G}$ and $\widetilde{\Sigma}$ obey the Schwinger-Dyson
equations \begin{eqnarray}
G_{\mathbf{q}}^{r} & = & G_{0\mathbf{q}}^{r}+G_{0\mathbf{q}}^{r}\otimes\Sigma_{\mathbf{q}}^{r}\otimes G_{\mathbf{q}}^{r}\label{SDr}\\
C_{\mathbf{q}} & = & G_{\mathbf{q}}^{r}\otimes\left[G_{0\mathbf{q}}^{r-1}\otimes C_{0\mathbf{q}}\otimes G_{0\mathbf{q}}^{a-1}+\Sigma_{\mathbf{q}}\right]\otimes G_{\mathbf{q}}^{a}\,.\label{SDK}\end{eqnarray}
 We use $\otimes$ to distinguish the matrix product over time (the
time convolution) where $A\otimes B\,(t,t')\equiv\int dsA(t,s)B\left(s,t'\right)$
from the scalar (element-wise) product, where $A\, B\left(t,t'\right)=A\left(t,t'\right)B\left(t,t'\right)$.

The first Schwinger-Dyson equation \ref{SDr} is similar to conventional
perturbation theory. However it is coupled to the second equation
of $C\left(t_{1},t_{2}\right)$ which admits non-trivial solutions.
For simplicity, we choose an initial condition such that $C_{0}=0$
and $\texttt{$\textrm{Im}$}G_{0}^{r}=0$. This leads to thermalization
of $G^{r}$ (fulfillment of the FDT) in the absence of non-linearities,
so it is an appropriate condition. Let us call $t_{2}=t_{w}$ and
$t_{1}=t_{w}+\tau$, i.e., the correlations between instants separated
by $\tau$ are measured after the waiting time $t_{w}$ has elapsed.
The glassy dynamics appears in a regime of $t_{w}\rightarrow\infty$
and $\tau\rightarrow\infty$ . To proceed we assume that in this limit
all correlation functions can be decomposed into a slow, non-time
translation invariant, aging part ($AG$) and a fast, time translation
invariant, stationary part ($ST$) like in Eq.\ref{splitagst}.

In this approach the stationary term decays on a characteristic time
scale $\tau_{\beta}$ and it represents the correlations between degrees
of freedom which are in equilibrium with the thermal reservoir at
temperature $T$, i.e, retarded and Keldysh correlation functions
are related by the FDT. The aging part on the other hand depends weakly
on $t_{w}$ (aging phenomena) and varies slowly on $\tau$ in a characteristic
large time scale $\tau_{\alpha}\gg\tau_{\beta}$, which allows us
to neglect its $\tau$ derivatives. Moreover, as in reference \onlinecite{CL99},
we enforce a relationship between $G_{\text{AG}}^{r}$ and $C_{\text{AG}}$
by defining an effective temperature $T_{\text{eff}}$ at which the
long time correlations thermalize through a generalized FDT relation
\begin{equation}
G_{\text{AG}}^{r}\left(\tau,t\right)=-\frac{1}{T_{\text{eff}}\left(t\right)}\frac{\partial}{\partial t}C_{\text{AG}}\left(\tau,t\right)\,.\label{GeneralFDT}\end{equation}
 For $t\ll\tau_{\alpha}$ we can neglect the time dependence of $T_{\text{eff}}$,
leading to a constant $T_{\text{eff}}\left(t\right)\rightarrow T_{\text{eff}}$.

Using the above assumptions, the first Schwinger-Dyson equation (\ref{SDr})
in the stationary regime can be solved by a Fourier transform, which
gives\begin{equation}
\mathcal{G}_{\text{ST}\mathbf{q}}^{r}\left(\omega\right)=\frac{1}{\mathcal{G}_{0\mathbf{q}}^{r-1}\left(\omega\right)-\Sigma_{\text{ST}}^{r}\left(\omega\right)}\,.\label{SDrst}\end{equation}
 Here we will explore solutions where $\,\, G_{\text{AG}}^{r}\left(\tau,t_{w}\right)$
is small but finite. This corresponds to the weak long term scenario
where, even though $\lim_{\tau\rightarrow\infty}\lim_{t\rightarrow\infty}G_{\text{AG}}^{r}\left(\tau,t\right)\rightarrow0$,
the integral $\lim_{t\rightarrow\infty}\int_{0}^{t}dt^{\prime}G_{\text{AG}}^{r}\left(\tau,t^{\prime}\right)$
is still finite. In other words, we consider that the system keeps
a vanishingly small memory of what happened in the past which, when
accumulated over long times, gives a finite contribution to the dynamics.
Thus, following along the lines of Ref.\onlinecite{CL99}, we get
for the for the aging regime

\begin{equation}
G_{\text{AG}\,\mathbf{q}}^{r}\left(\tau,t_{w}\right)=\Sigma_{\text{AG}}^{r}\left(\tau,t_{w}\right)\mathcal{G}_{\text{ST}\mathbf{\, q}}^{r}\left(0\right)^{2}\,.\label{SDrag}\end{equation}

Analogously, the second Schwinger-Dyson equation (\ref{SDK}) can
be solved for the same two regimes, yielding\begin{equation}
G_{\text{ST}\mathbf{\, q}}^{K}\left(\omega\right)=\Sigma_{\text{ST}}^{K}\left(\omega\right)\mathcal{G}_{\text{ST}\mathbf{\, q}}^{r}\left(\omega\right)^{2}\label{SDKst}\end{equation}
 and\begin{equation}
C_{\text{AG}\,\mathbf{q}}\left(\tau,t_{w}\right)=\frac{\Sigma_{\text{AG}}\left(\tau,t_{w}\right)\mathcal{G}_{\text{ST}\mathbf{\, q}}^{r}}{\mathcal{G}_{\text{ST}\mathbf{\, q}}^{r}\left(0\right)^{-1}-\frac{1}{T_{\text{eff}}}\Sigma_{\text{AG}}^{K}\left(\tau,t_{w}\right)}\,.\label{SDKag}\end{equation}
 Note that if we identify the real time correlation functions with
the inter-replica correlation functions $\mathcal{F}$ and $\mathcal{G}$
as follows\begin{eqnarray}
\lim_{\tau\rightarrow\infty}\lim_{t_{w}\rightarrow\infty}C_{\text{AG}}\left(\tau,t_{w}\right) & \equiv & \mathcal{F}\label{r1}\\
\mathcal{G}_{\text{ST}}^{r}\left(\omega\right) & \equiv & \mathcal{G}\left(i\omega_{n}\rightarrow\omega+i\delta\right)\label{r3}\end{eqnarray}
 we get that the Schwinger-Dyson equations \ref{SDrst} and \ref{SDKag}
are exactly equivalent to the replica equations in \ref{D}, provided
that the self energies in the two schemes are also equivalent, i.e.,
$\Sigma_{\text{AG}}=-\Sigma_{\mathcal{F}}$ and $\Sigma_{\text{ST}}^{r}\left(\omega\right)=\Sigma_{\mathcal{G}}\left(i\omega_{n}\rightarrow\omega+i\delta\right)$.
To prove this last requirement, we apply the same one-loop self-consistent
screening perturbative scheme for the real time DMFT self energies
$\widetilde{\Sigma}$. This gives\[
\widetilde{\Sigma}=\left(\begin{array}{cc}
0 & C\otimes D^{a}+D\otimes G^{a}\\
C\otimes D^{r}+D\otimes G^{r} & \,\, D^{a}\otimes G^{a}+D^{r}\otimes G^{r}+D\otimes C\end{array}\right)\]
 where dressed interactions are given by\begin{eqnarray}
D^{r} & = & u+u\Pi^{r}\otimes D^{r}\label{DIr}\\
D & = & u\Pi^{r}\otimes D+u\Pi\otimes D^{a}\,,\label{DIK}\end{eqnarray}
 and the polarization bubbles are given by the element-wise products
(no time or space convolutions) \begin{eqnarray}
\Pi^{r} & = & \gamma C\, G^{r}\label{Pr}\\
\Pi & = & \gamma\left[\left(G^{a}\right)^{2}+\left(G^{r}\right)^{2}+C^{2}\right]\,.\label{PK}\end{eqnarray}
 The stationary and aging contributions to the self energies and polarizations
can be calculated analogously by using the definitions (\ref{Pr})
and (\ref{PK}) and separating the stationary from the aging contribution
according to their asymptotic time behavior. This gives\begin{eqnarray}
\Sigma_{\text{ST}}^{r} & = & C_{\text{AG}}\otimes D_{\text{ST}}^{r}+D_{\text{AG}}\otimes G_{\text{ST}}^{r}\label{Sr}\\
 &  & +C_{\text{ST}}\otimes D_{\text{ST}}^{r}+D_{\text{ST}}\otimes G_{\text{ST}}^{r}\nonumber \\
\Sigma_{\text{AG}} & = & D_{\text{AG}}\otimes G_{\text{AG}}\label{Sk}\end{eqnarray}
 and\begin{eqnarray}
\Pi_{\text{ST}}^{r} & = & \gamma\left(C_{\text{AG}}G_{\text{ST}}^{r}+C_{\text{ST}}G_{\text{ST}}^{r}\right)\label{Pirag}\end{eqnarray}
 and \begin{eqnarray}
\Pi_{\text{AG}} & = & \gamma C_{\text{AG}}^{2}\label{PiKag}\end{eqnarray}
 It is easy to verify accordingly that $\Pi_{\text{ST}}$ and $\Pi_{\text{ST}}^{r}$
are related by FDT at the temperature $T$ and $\Pi_{\text{AG}}$
and $\Pi_{\text{AG}}^{r}$ are related by the classical FDT at a temperature
$T_{\text{eff}}$ . The last term in Eq.\ref{Sr} is exactly the Matsubara
convolution on (\ref{scsa1}). The remaining terms are calculated
as follows:\begin{equation}
\mathcal{D}_{\text{ST}}^{r}\left(\omega\right)=\frac{1}{u^{-1}+\Pi_{\text{ST}}^{r}\left(\omega\right)}\label{DIrst}\end{equation}
\begin{equation}
D_{\text{AG}}=\frac{-\Pi_{\text{AG}}\mathcal{D}_{\text{ST}}^{r}\left(\omega=0\right)}{\mathcal{D}_{\text{ST}}^{r}\left(\omega=0\right)^{-1}+\frac{1}{T_{\text{eff}}}\Pi_{\text{AG}}}\label{DIKag}\end{equation}
 Using the relations \ref{r1} and \ref{r3} and performing the usual
Matsubara sums we get,$\Pi_{\mathcal{F}}=\Pi_{\text{AG}}$ and $\Pi_{\mathcal{G}}\left(i\omega_{n}\rightarrow\omega+i0\right)=\Pi_{\text{ST}}^{r}\left(\omega\right)$
which yields $D_{\text{AG}}=D_{\mathcal{F}}$ and$D_{\mathcal{G}}\left(i\omega_{n}\rightarrow\omega+i0\right)=\mathcal{D}_{\text{ST}}^{r}\left(\omega\right)$.
Therefore, equations (\ref{DIrst}), (\ref{DIKag}), (\ref{Sr}),
and (\ref{Sk}) together prove that $\Sigma_{\text{AG}}=\Sigma_{\mathcal{F}}$
and $\Sigma_{\text{ST}}^{r}\left(\omega\right)=\Sigma_{\mathcal{G}}\left(i\omega_{n}\rightarrow\omega+i\delta\right)$and
that there is a complete connection between replica and Schwinger-Keldysh
formalisms. There is still one more independent equation, namely,
Eq. \ref{SDrag}, in the Schwinger-Keldysh formalism which bares no
analog on the Dyson equations for the inter-replica correlation functions.
However, if we integrate over $\mathbf{q}$ both sides of \ref{SDrag},
substitute the definition $\Sigma_{\text{AG}}^{r}=C_{\text{AG}}\otimes D_{\text{AG}}^{r}+D_{\text{AG}}\otimes G_{\text{AG}}^{r}$
into \ref{SDrag} with $D_{\text{AG}}^{r}=\Pi_{\text{AG}}^{r}\mathcal{D}_{\text{ST}}^{2}\left(0\right)$we
obtain\begin{equation}
\left[2\mathcal{D}_{\text{ST}}^{2}\left(0\right)\Pi_{\text{AG}}^{K}-D_{\text{AG}}^{K}\right]\int\frac{d^{d}q}{\left(2\pi\right)^{d}}\mathcal{G_{\text{ST}}}_{\mathbf{q}}^{2}\left(0\right)=1\label{margfinal}\end{equation}
 which, is exactly the marginality condition of the replica approach.
This proves our previous statement that the saddle point condition
gives the dynamical behavior at the time scales $t_{\mathrm{w}}\rightarrow\infty$,
$\tau\rightarrow\infty$ with $t_{\mathrm{w}}\ll\tau_{\alpha}$. In
this limit the effective temperature is not yet in equilibrium. In
the opposite limit one cannot disregard the time dependence of $T_{\text{eff}}\left(t\right)$and
it is thus not possible to write a closed form like equation (\ref{margfinal}).
Nevertheless, the equilibrium approach based on the replica trick,
together with the droplet relaxation picture enables us to access
the behavior of $T_{\text{eff}}\left(t_{\text{w}}\right)$in the time
limit $t_{\mathrm{\text{w}}}\gg\tau_{\alpha}$.

\end{document}